\documentstyle[aps,preprint,psfig]{revtex}
\tighten
\begin{document}

\preprint{LBNL--41395}

\title{
Medium Modified Cross Sections, Temperature and Finite Momentum Effects for 
Antikaon Production in Heavy-Ion Collisions%
\thanks{This work was supported by a NATO collaborative research grant number 
970102. This work was also partly supported by BMBF, DFG, GSI, Alexander von
Humboldt-Stiftung, and  
by the Director, Office of Energy Research,
Office of High Energy and Nuclear Physics,
Nuclear Physics Division of the U.S. Department of Energy
under Contract No.\ DE-AC03-76SF00098.}}

\author{J\"urgen Schaffner-Bielich}

\address{RIKEN BNL Research Center, Brookhaven National Laboratory,\\
Upton, New York 11973, USA}

\author{Volker Koch}

\address{Nuclear Science Division, Lawrence Berkeley National Laboratory,\\
University of California, Berkeley, California 94720, USA}

\author{Martin Effenberger}

\address{Institut f\"ur Theoretische Physik, Universit\"at Giessen, 
D--35392 Giessen, Germany}

\date{\today}
\maketitle

\begin{abstract}
  The medium modifications of antikaons in dense matter are studied in a
  coupled channel calculation for scenarios more closely related to the
  environment encountered in heavy-ion collisions. We find that the optical
  potential of the antikaons turns repulsive or is drastically reduced for
  finite momenta or finite temperature. 
  Hence, the antikaon mass does not decrease substantially in
  heavy-ion collisions to provide an explanation for the observed antikaon
  production rates at threshold. We demonstrate that the in-medium production
  cross section of antikaons via pions and $\Sigma$ hyperons is remarkably
  enhanced up to an order of magnitude. The effect of in medium correction 
  on the resulting kaon spectra is studied within a transport model. 
  We find that both in medium correction lead to about the same enhancement 
  of the spectrum. However, once the temperature dependence is taken 
  into account no enhancement is found. 
\end{abstract}

\draft


\section{Introduction}

Antikaons have been proposed as a tool to probe in-medium effects
in heavy-ion collisions. The work of Nelson and Kaplan \cite{NK87}
speculating about a strange condensate realignment in relativistic heavy ion
collision by antikaons initiated numerous work in this field. Indeed, coupled
channel calculations for antikaons in matter \cite{Koch94} demonstrate that the
K$^-$ feels an attractive potential in 
dense matter, which was confirmed by 
\cite{Waas96b,Waas97} and \cite{Lutz98}. The main effect stems from the
$\Lambda(1405)$ resonance which is shifted upwards at finite density (see
\cite{Sta87} and references therein).
Also, data from kaonic atoms indicates that the K$^-$
feels a sizable attraction inside the nucleus \cite{Fried94}.
More theoretical foundation for a reduced effective antikaon mass 
has been also given using chiral perturbation theory
\cite{Brown94,Lee95} and mean-field models \cite{Maru94,Scha94b,SMB97}.

This has triggered a considerable interest in the production
of K$^-$ in heavy-ion collisions at threshold.
Indeed, an enhanced production of K$^-$ has been
observed at GSI \cite{Kaos97,Fopi96}.
This phenomenon has been attributed to a reduction of the antikaon
effective mass in the dense medium \cite{Li94,Li96,Li97,Cassing97}. 

An attractive potential for antikaons has been extracted from kaonic
atoms at zero momentum \cite{Fried94}. However, heavy-ion collisions probe
the optical potential of the antikaon at a finite momentum.
Assuming a temperature of $T=80$ MeV \cite{Fopi96}, the antikaon has an average
momentum of more than 300 MeV with respect to the matter restframe. 
Furthermore, all experimental data available so far cover only momenta above
$\sim 300 \,\rm MeV$ with respect to the matter rest frame. Calculations 
have so far assumed that the attractive potential for the
kaons i.e. their change in the mass is {\em independent} of momentum.
This constitutes a considerable extrapolation. To which extend this is
justified will be addressed in the present paper.

Based on a dynamical model for the K$^-$--nucleon interaction \cite{Koch94}, 
we will argue 
that the antikaon optical potential in the momentum range relevant for 
heavy-ion collisions is likely to be small if not  
repulsive. Only for the smallest momenta, which are considerably below
the range covered by experiment, we find attraction which is
compatible with fits to kaonic atoms. Including the selfenergy of the kaon in
a selfconsistent calculation we find that the optical potential at finite
momentum is attractive but considerably reduced.
Thus it is appears unlikely, that the reduction of the
antikaon mass is the underlying in-medium effect responsible for the enhanced
K$^-$ production rate. Instead, we find that the in-medium cross
section for producing antikaons via $\Sigma$ hyperons might be the relevant
mechanism for the observed enhancement. There are indications from stopped
K$^-$ data on nuclei that there is indeed a change of the branching ratios in
the nuclear medium \cite{Ohnishi97} supporting that picture.
Nevertheless, we also demonstrate that temperature effects are crucially
important in the sense that they can wash out any in-medium effect.

The paper is organized as follows: we introduce the coupled channel approach
in Sect.\ \ref{sec:coupled} and calculate medium effects for the kaon optical
potential and cross section in Sect.\ \ref{sec:coupled}. Effects of the nucleon 
and kaon selfenergy are studied in Sect.\ \ref{sec:self}. The in-medium effects 
for antikaons are included in a BUU transport model  in Sect.\ \ref{sec:production} and
contrasted with experimental data. 


\section{Coupled channel model for antikaons}
\label{sec:coupled}

In the following we adopt the nonperturbative coupled channel approach of
\cite{Koch94,Ohnishi97} starting from the coupled channel Schr\"odinger
equation,
\begin{equation}
\nabla^2 \psi_i(r) + k_i^2 \psi_i(r)
         - 2 \mu_i \int V_{i,j}(r,r') \, \psi_j(r') d^3r' = 0\ ,
\label{eq:Sch}
\end{equation} 
where $\psi_i(r)$ represents the relative wave function, $\mu_i$
is the reduced mass, and $k_i$ is the momentum in the center of mass system
for the corresponding channel. For a separable potential of the form 
\begin{equation}
V_{i,j}(k,k') =  g^2 C_{i,j} \,\, v_i(k) \, v_j(k')\\
              = \frac{g^2}{\Lambda^2} C_{i,j} \Theta(\Lambda^2 - k^2 )
                  \Theta(\Lambda^2 - k'^2 )
\label{eq:seppot}
\end{equation}
the scattering $T$-matrix simplifies to
\begin{equation}
\label{eq:Tmat}
T_{i,j}(k,k',E) = g^2 \,v_i(k) \, v_j(k')
        \left[ (1- C \cdot G(E))^{-1} \cdot C \right]_{i,j}\ .
\end{equation}
The propagator matrix $G$ is diagonal and in the vacuum given by
\begin{eqnarray}
g_i(E) &=&  g^2 \,
          \int \frac{d^3 p}{(2 \pi)^3} 
                \frac{v_i^2(p)}{E - m_i - M_i  - p^2/(2 \mu_i)} \\
        &=&\frac{1}{2 \pi^2} \frac{g^2}{\Lambda^2}
                \int_0^\Lambda
                        \frac{p^2 \, dp}{E - m_i - M_i - p^2/(2 \mu_i)}\ .
\label{eq:prop}
\end{eqnarray} 
The coupling matrix $C_{ij}$ is chosen to be of the standard
$SU(3)$ flavor symmetric form.  In addition to the previous approaches of
\cite{Koch94,Ohnishi97}, we include also the isospin one channels explicitly.
For isospin zero, there are two channels
\begin{equation}
\label{eq:couple}
C^{I=0}_{ij} =
\begin{array}{cc}
\begin{array}{cc}
\bar{K}N & \pi \Sigma
\end{array} & \\
\left(\begin{array}{cc}
-\frac{3}{2} & -\frac{\sqrt{6}}{4}\\
\\
-\frac{\sqrt{6}}{4} & -2
\end{array}\right) &
\begin{array}{c} \bar{K}N\\ \\ \pi \Sigma \end{array}
\end{array}
\quad .
\end{equation} 
whereas for the isospin one part, one has three channels which are coupled as
\begin{equation}
\label{eq:couple2}
C^{I=1}_{ij} =
\begin{array}{cc}
\begin{array}{ccc}
\bar{K}N & \pi \Sigma & \pi \Lambda 
\end{array} & \\
\left(\begin{array}{ccc}
-\frac{1}{2} &-\frac{1}{2} & -\frac{\sqrt{6}}{4}\\
\\
-\frac{1}{2} &-1 & 0 \\
\\
-\frac{\sqrt{6}}{4} & 0 & 0
\end{array}\right) &
\begin{array}{c} \bar{K}N\\ \\ \pi \Sigma\\ \\ \pi \Lambda \end{array}
\end{array}
\quad .
\end{equation} 
The coupled channel calculation is performed for the two isospin channels
separately. 
The scattering amplitude in a given isospin channel is then
\begin{equation}
\label{eq:ampl}
f^I_{i,j}(k,k') = - \frac{\sqrt{\mu_i\mu_j}}{2 \pi} T^I_{i,j}(k,k')\ .
\end{equation}
The cross section of a given channel is given by
\begin{equation}
\sigma_{ij} = 4 \pi \frac{k'}{k} |f_{ij}(k,k')|^2 
\label{eq:cross}
\end{equation}
and is constructed out of the two isospin components of the $T$ matrix.

The two parameters $\Lambda$ and $g$ can be fixed either by 
the scattering amplitude as extracted by Martin \cite{Martin81}
or by the threshold ratios as in \cite{Siegel88}
\begin{eqnarray}
\gamma &=& \frac{\Gamma(K^-p\to\pi^+\Sigma^-)}{\Gamma(K^-p\to\pi^-\Sigma^+)}  
= 2.36 \pm 0.04 \\
R_c &=& \frac{\Gamma(K^-p\to\mbox{charged
particles})}{\Gamma(K^-p\to\mbox{all}) } = 0.664 \pm 0.011 \\
R_n &=& \frac{\Gamma(K^-p\to\pi^0\Lambda)}{\Gamma(K^-p\to\mbox{all neutral
states})} = 0.189 \pm 0.015 \quad .
\end{eqnarray}
We choose the former way as in \cite{Koch94} and take over the parameters  
$\Lambda=0.78$ GeV and $g^2=17.9$.  
We note in passing, that this choice of parameters gives the threshold ratios 
$\gamma=4.11$, $R_c=0.627$, and $R_n=0.234$. With a slight change of the
coupling constant, i.e.\ for $g^2=16.8$, one gets the values
$\gamma=2.37$, $R_c=0.644$, and $R_n=0.121$ but the pole position of the
$\Lambda(1405)$ is shifted then towards 1420 MeV. This is consistent with the
findings in \cite{Siegel88}, that one can either describe the scattering
amplitude or the threshold ratios with a slight change of the coupling
constant. 
We have verified that our results presented in the following 
are not sensitive to these small changes in the 
coupling constant $g$.  

Nuclear medium effects can be easily included. Pauli blocking effects forbid
intermediate nucleon states with a momentum less than the Fermi momentum.
Hence, the propagator in the $\bar{K}N$ channel is modified in the medium to
\begin{equation}
\label{eq:propmat}
g_1(E,k_f) = \frac{g^2}{\Lambda^2}
        \int_0^\Lambda \frac{d^3 p}{(2\pi)^3} \,
        \frac{\Theta(k_f - | \vec{p}+M_N\vec{v}_\Lambda |)}
                {E - m_K - M_N - p^2/(2 \mu_{Kp})}
\end{equation}
in both isospin channels. 
Here, effects from a moving $\Lambda(1405)$ resonance is taken into account
by the term proportional to the velocity of the resonance $v_\Lambda$
\cite{Ohnishi97}. 
For a kaon with finite momentum with respect to the nuclear matter rest frame
the intermediate nucleon propagator will be shifted in momentum
space. Hence, the Pauli blocking effect will be reduced in that case and the
effects stemming from the $\Lambda(1405)$ resonance are moderated. 

The possibility for large range terms in the kaon-nucleon interaction, which
might be important in the medium, has been raised in \cite{Lee95}.
Fits to the K$^-$-N scattering data using a coupled channel formalism 
find a good description without a range term \cite{Kaiser95,Oset98}. 
Also in the model presented above (without range terms) one gets a good
agreement with the scattering amplitude far below threshold \cite{Koch94}. 
Hence, we will not consider range terms in the following.


\section{Optical potential and in-medium cross sections}
\label{sec:optical}

\begin{figure}[htbp]
\begin{center}
\leavevmode
\psfig{file=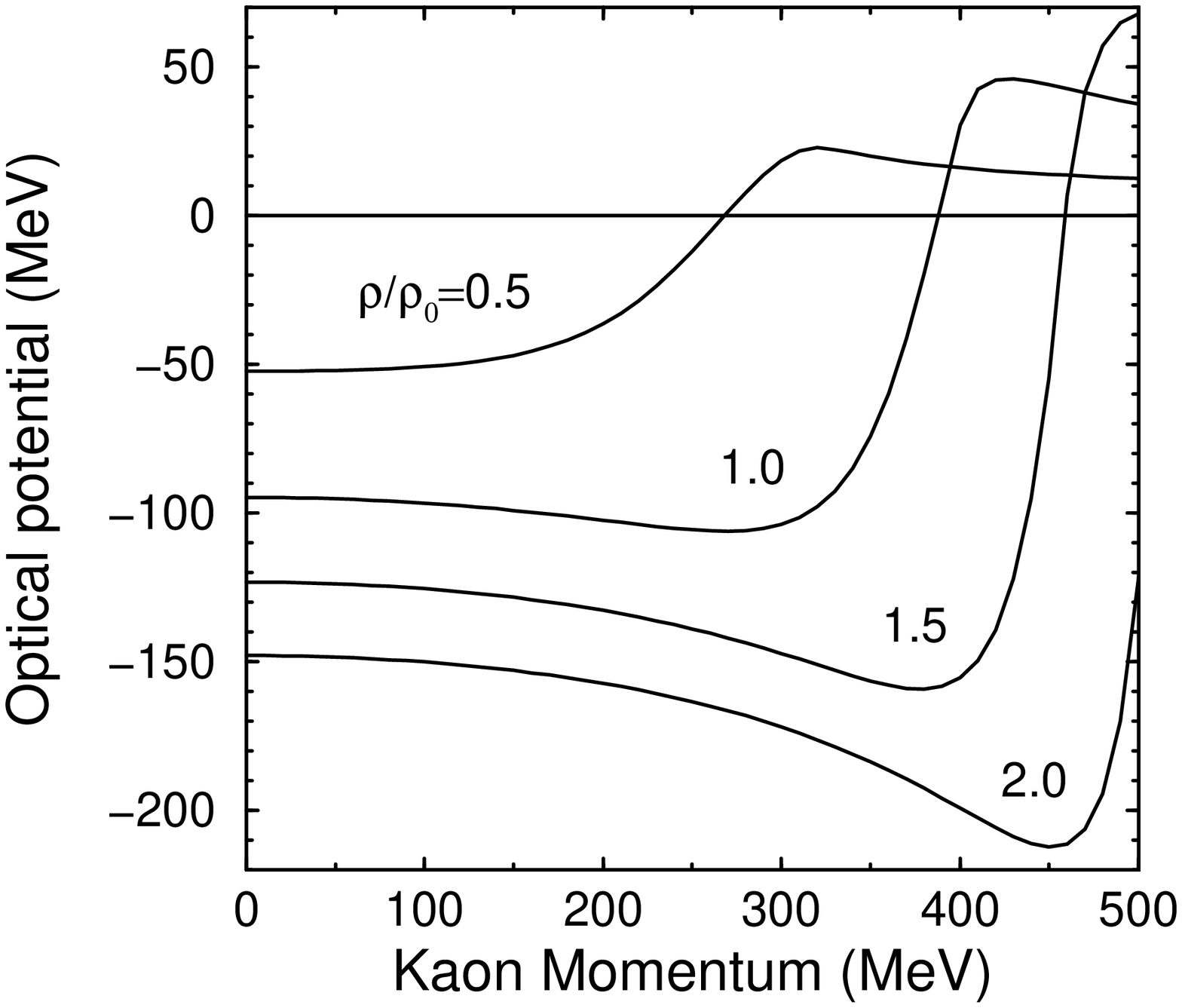,height=0.35\textheight} 
\end{center}
\caption{The antikaon optical potential versus the antikaon
relative momentum for different densities. 
The optical potential turns repulsive at a certain
momentum especially for lower densities.
At densities above $2\rho_0$, the optical potential  stays attractive even for
high relative momenta.} 
\label{fig:uopt_pkaon}
\end{figure}

Antikaons produced in heavy-ion collisions will change their properties due to
the surrounding nucleons. The antikaons produced in the first collisions
will suffer rescattering processes from the in-streaming nucleons.
The surrounding matter is not like bulk matter, it is more like a streaming
matter with a high momentum. Therefore, the optical potential relevant for the
production of kaons even at subthreshold energies has to be considered at a
finite momentum in the range of $p\approx 250$ MeV relative to the matter rest 
frame. In the following we denote this shift in momentum space of the
colliding antikaon-nucleon pair (with a given colliding energy $\sqrt{s}$) 
with respect to the rest frame of the nuclear
bulk matter in short form as the {\em relative momentum}.
Figure \ref{fig:uopt_pkaon}
shows the optical potential of the kaon as defined as
\begin{equation}
U_{\mbox{opt}} = - \frac{2\pi}{\mu_{Kp}} 
\Re \left( \frac{1}{4}f^{I=0}_{11}+\frac{3}{4}f^{I=1}_{11} \right) \rho
\label{eq:uopt}
\end{equation}
where we ignore recoil corrections. They are small \cite{Koch94} and 
not relevant for the implementation into cascade models.
For low relative momenta, the optical potential is getting more and more attractive as
the density increases. This changes remarkably at higher relative momentum. The optical
potential changes sign and gets repulsive again. Especially for moderate
densities, at normal nuclear density $\rho_0$ or less, the change happens at
relative momenta not far from the Fermi momentum of the nucleons. For higher
relative momenta, 
the optical potential stays repulsive. Hence, antikaon production is less
favored at high relative momenta, on the contrary it  
is even slightly suppressed in the medium.
Pauli blocking effects of the propagator in the $\bar
KN$ channel, eq.\ (\ref{eq:propmat}), are responsible for switching the
sign of the optical potential from repulsion at low density to attraction at
high density (this is due to the $\Lambda(1405)$, see \cite{Koch94}).  This
mechanism is not working at high relative momenta 
(comparable to the Fermi momenta of the nucleons), therefore resulting in a
repulsive optical potential for the antikaons. For a higher density of
$2\rho_0$ and more the $\Lambda(1405)$ melts in the medium and an attractive
potential is seen even for large relative momenta. Therefore, one has to
focus either on high density or on low to moderate density with small relative 
momentum to see the in-medium reduction of the antikaon mass in
matter. Inclusive observables in heavy-ion collisions are likely to probe
regions of phase space where the antikaon feels both attractive and repulsive potentials
so that in-medium effects are washed out. We will discuss these effects by
making a detailed comparison with experiment in section \ref{sec:production}.

\begin{figure}[htbp]
\begin{center}
\leavevmode
\psfig{file=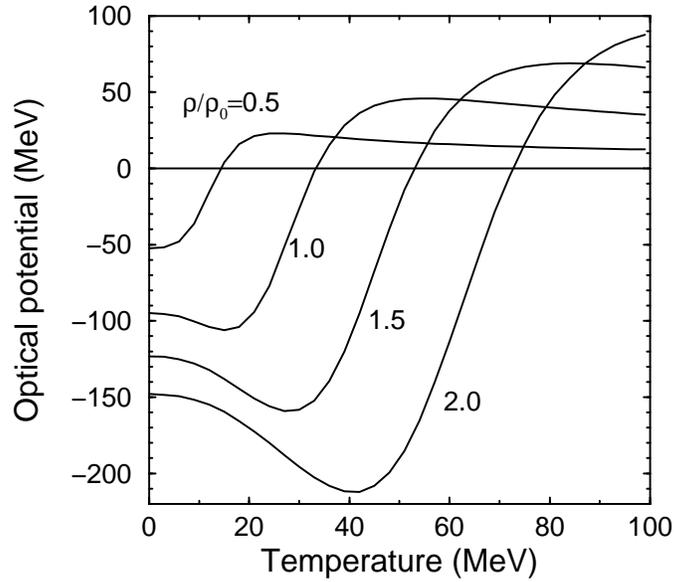,height=0.35\textheight} 
\end{center}
\caption{The antikaon optical potential as a function of the temperature of the
nucleons for different densities. The optical potential turns
repulsive at a rather low temperature.}
\label{fig:uopt_temp}
\end{figure}

In the intermediate and final stage of the heavy-ion collision, the nuclear
matter will be heated up. Typical slope parameters are in the range of $T=90$
MeV for the kaons at a bombarding energy of 1.8 AGeV \cite{Kaos97}.
Temperature effects can be implemented in our approach by replacing the
$\Theta$ function in the propagator (\ref{eq:propmat}) with a
Fermi-Distribution:
\begin{equation}
\label{eq:proptemp}
g_1(E,k_f) = \frac{g^2}{\Lambda^2}
        \int_0^\Lambda \frac{d^3 p}{(2\pi)^3} \,
        \frac{\left[{\rm e}\,^{(p^2/(2M_N)-\nu)/T} + 1 \right]^{-1}}
{E - m_K - M_N - p^2/(2 \mu_{Kp})}
\quad ,
\end{equation}
where $\nu$ is the chemical potential which is fixed for a given density
$\rho$. Again, Pauli blocking effects will be weakened by a thermal smearing of
the Fermi sphere. Indeed, as seen in Fig.\ \ref{fig:uopt_temp}, the optical
potential for antikaons changes considerably for a finite temperature.
At a temperature of $T=90$ MeV, the optical potential has turned from
attraction to repulsion for all cases shown up to a density of
$\rho=2\rho_0$. The temperature at which the change of signs happens
is rather low, especially for moderate densities. At normal nuclear density,
this critical temperature is around $T=30$ MeV.

Usually, the optical potential
for $T=0$ MeV is used for describing the enhanced production of antikaons in
heavy-ion collisions. The optical potential is repulsive at very low densities
up to $\rho\approx 0.2\rho_0$ due to the $\Lambda(1405)$ resonance. As the
resonance mass is shifted up in dense matter (see \cite{Sta87}), the optical
potential of the antikaons turns attractive for higher density. 
The switch in the sign of the optical potential from attraction to repulsion
happens to occur at 
a much higher density, if the surrounding matter is moderately heated up. 
For a temperature of $T=90$ MeV, this turning point is already
shifted beyond $2\rho_0$.

\begin{figure}[htbp]
\begin{center}
\leavevmode
\psfig{file=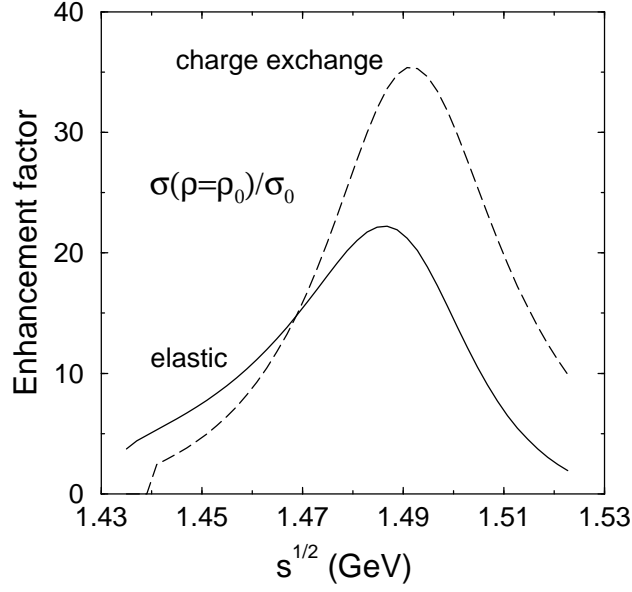,height=0.35\textheight} 
\end{center}
\caption{The enhancement factor of the charge exchange reaction
$K^-+p\to K^0+n$ and the elastic reaction at normal nuclear matter density.}
\label{fig:cross_enh2}
\end{figure}

While the optical potential is essentially zero for momenta relevant
for heavy ion collisions, we find a considerable increase of the
production cross section $\pi + \Sigma \rightarrow K^- + N$. 
Figure
\ref{fig:cross_enh2} shows this enhancement factor which we define as
the ratio of the cross section at normal nuclear density and the vacuum cross
section $\sigma(\rho=\rho_0)/\sigma_0$ for the elastic reaction $K^-p\to K^-p$
and the charge exchange reaction $K^-p\to K^0n$. Both cross sections are
enhanced tremendously. The elastic cross section is four times bigger already
at threshold and increases up to a factor of 27 around an antikaon momentum
close to the Fermi momentum of the nucleons. The charge exchange reaction sets
in at a finite antikaon ($K^-$) momentum due to the mass difference of the
charged and neutral antikaon. This cross section is even more enhanced in the
medium than the elastic cross section signaling significant rescattering
effects for the production of $K^-$ in the medium. 

\begin{figure}[htbp]
\begin{center}
\leavevmode
\centerline{\psfig{file=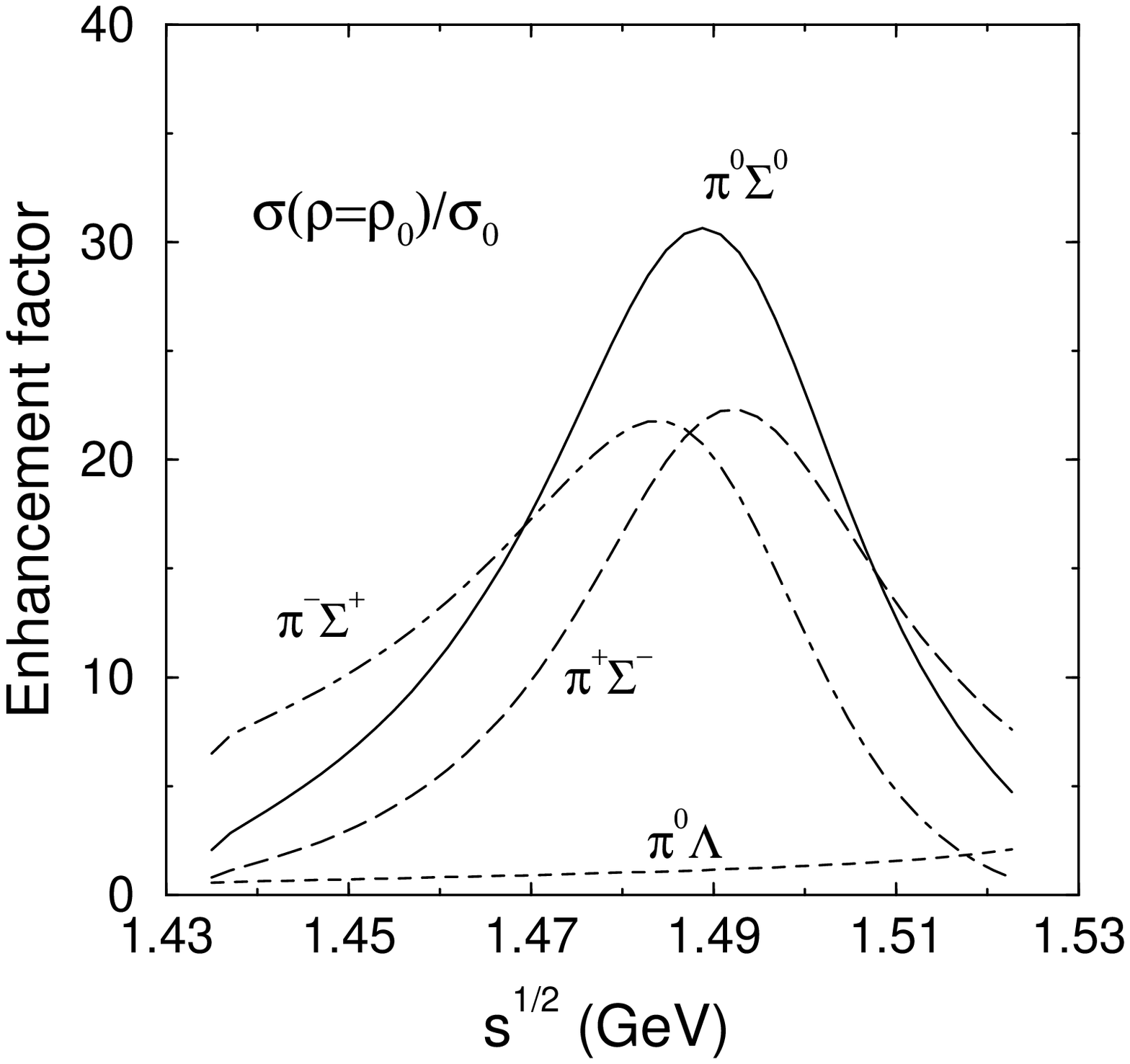,height=0.35\textheight} 
\psfig{file=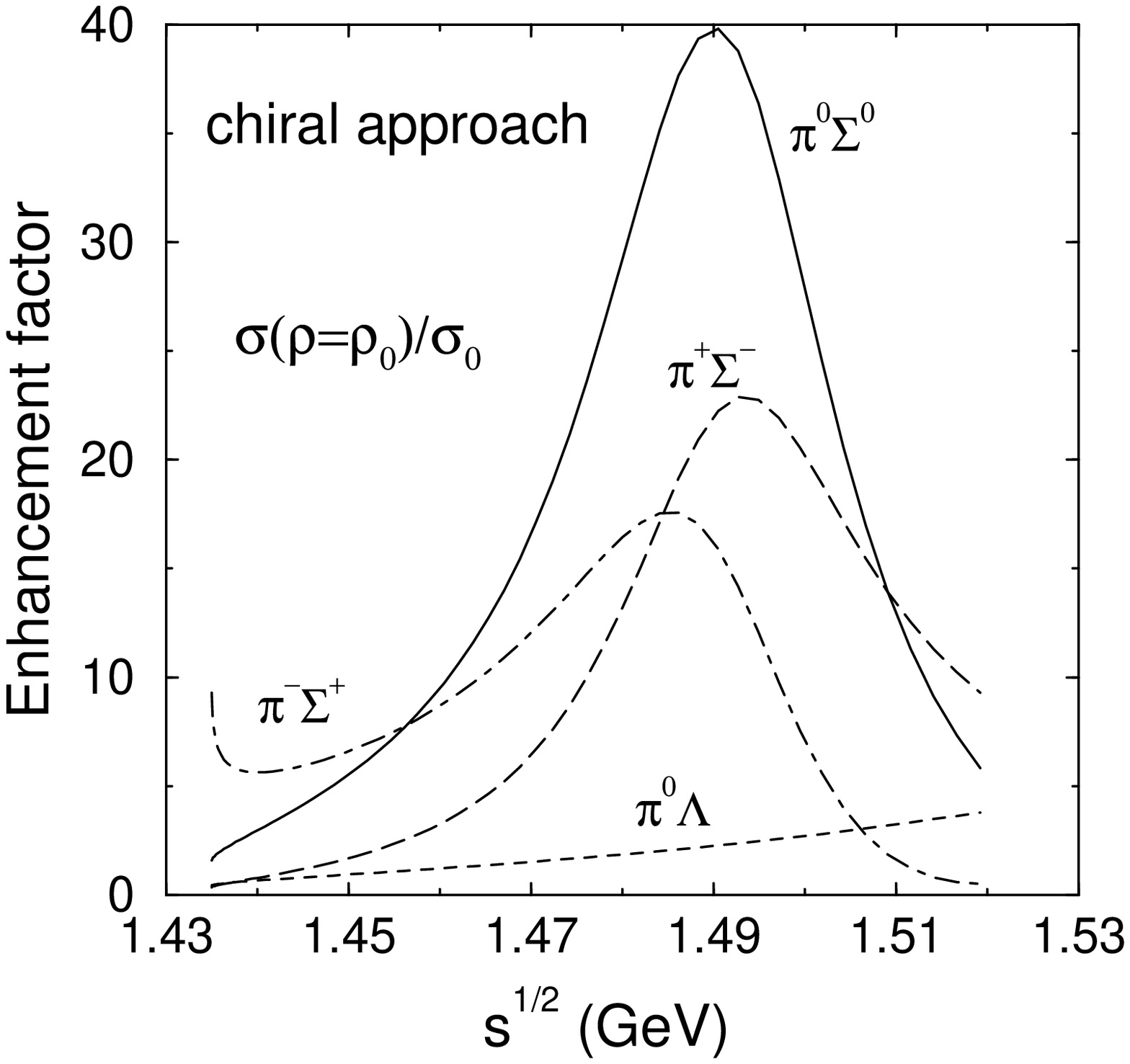,height=0.35\textheight}}
\end{center}
\caption{Left figure: The enhancement factor of the antikaon production cross sections at
normal nuclear density. The process $\pi+\Sigma\to p+K^-$ is considerably
enhanced in the medium.\\
Right figure: The same as left but using 
the nonperturbative chiral approach of
\protect\cite{Oset98} with relativistic propagators.}
\label{fig:cross_enh}
\label{fig:cross_enh_rel}
\end{figure}

Via detailed balance and by using eq.\ (\ref{eq:cross}) we calculate the
production cross section of antikaons in the medium by $\pi Y\to K^-p$.
We expect that especially the channels involving $\Sigma$ hyperons will be
enhanced in the medium as this channel enters into the isospin zero coupled
channel. The main medium effects are resulting from the dynamics of the
$\Lambda(1405)$ which changes the isospin zero part of the coupled channel
calculation. The channel $\pi \Lambda\to K^-p$, which is isospin one, is then not changed
significantly as evident from Fig.\ \ref{fig:cross_enh}. We note that
the antikaon production cross section $\pi\Sigma\to K^-p$ has increased by a
large factor in matter. Especially, the reaction $\pi^-\Sigma^+\to K^-p$ is
favored in dense matter compared to the other reactions at threshold. This
results in a change of the branching ratio for nuclear targets, as discussed in
\cite{Ohnishi97}. We point out, that the present experimental data seems to
support this medium effect for the reaction $(K^-,K^+)$ (see
\cite{Ohnishi97}). For higher antikaon momentum around 250 MeV, all three
$\Sigma$ channels are enhanced by a factor of $20-30$. 

The results presented here are not an artifact of our model used. One could
argue that a chiral nonperturbative approach might alter the picture. Also the
$\eta Y$ channel contributions have to be taken into account, so that the
matrices (\ref{eq:couple}) and (\ref{eq:couple2}) have to extended
correspondingly. Moreover, especially concerning the pion dynamics,
relativistic effects can be important. We have checked for all of these effects
by using the recent chiral nonperturbative approach of Oset and Ramos
\cite{Oset98} which uses relativistic propagators and includes effects from the
$\eta Y$ channels (for a recent extension of their model see \cite{Ramos99}). 
The propagators are solved numerically by using a Cauchy
integrator. The separable potential originating from the lowest order chiral
Lagrangian term are now of the form
\begin{equation}
V_{ij} = -C_{ij} \frac{1}{4 f^2} \left(k^0+k^{'0}\right)
\end{equation}
and momentum dependent. The calculation of the $T$-matrix in the medium
follows essential along the lines as described in section \ref{sec:coupled}.
The coupling matrix 
as well as the expression for the propagator and the cross section can be found
in \cite{Oset98}. 
Note that this chiral
approach provides a good description of the available vacuum scattering data.
In the following, we take the parameters of \cite{Oset98}, $f=1.15f_\pi$ and a
cutoff of $0.63$ GeV.
We remark on passing that the threshold ratios in the isospin
basis seems to be off but can be adjusted to the experimental data by a small
change of the parameters. 
For a coupling constant of $f=122.9$ MeV and a
momentum cutoff of $\Lambda=0.8$ GeV, we find the threshold ratios
$\gamma=2.35$, $R_c=0.635$, and $R_n=0.178$ which are in agreement with the
data except for $R_c$ which turns out to be a little bit too small. 
As found for the non-relativistic case before, the position of the 
$\Lambda(1405)$ is shifted up towards 1.42 GeV.

Medium effects are introduced in a similar fashion as discussed before.
The resulting in-medium enhancement factor of the cross
section is shown in Fig.\ \ref{fig:cross_enh_rel} in comparison to the
nonrelativistic case.  The $\pi\Lambda$ channel is
not as enhanced as the $\pi\Sigma$ channels, but shows a rise in the chiral
approach compared to the nonrelativistic approach.  The $\pi^-\Sigma^+$
production channel is enhanced at threshold even for this approach. Also, all
three $\pi\Sigma$ channels are drastically enhanced around an antikaon momentum
of 250 MeV by a factor between 20--40.  These findings support the picture
already seen in the nonrelativistic case.  

\begin{figure}[htbp]
\begin{center}
\leavevmode
\psfig{file=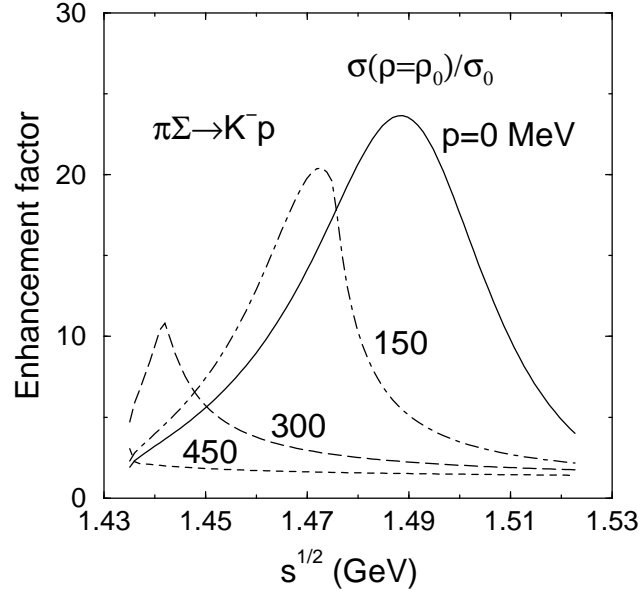,height=0.35\textheight} 
\end{center}
\caption{The hyperon induced cross section for antikaon production at finite
relative momenta with respect to the matter rest frame.} 
\label{fig:cross_nr_pkaon}
\end{figure}

\begin{figure}[htbp]
\begin{center}
\leavevmode
\psfig{file=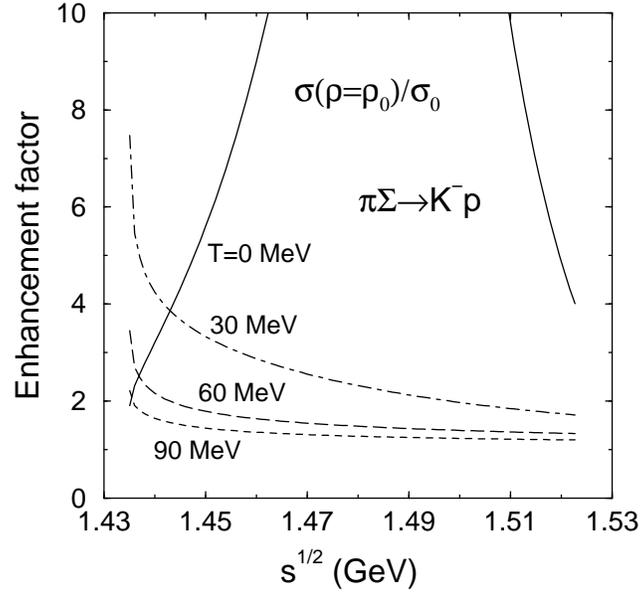,height=0.35\textheight} 
\end{center}
\caption{The hyperon induced cross section for antikaon production at finite
temperature.} 
\label{fig:cross_nr_temp2}
\end{figure}

Now, we turn to study effects of a finite relative momentum with respect to
the matter rest frame and temperature in 
the nonrelativistic case.
The enhancement factor for
the production of K$^-$ are depicted in Fig.\ \ref{fig:cross_nr_pkaon} at
finite relative momentum and in Fig.\ \ref{fig:cross_nr_temp2} for finite temperature,
respectively. Both plots show the average cross section for producing a K$^-$
via $\Sigma$ hyperons. Similar to the optical potential the effect decreases at
finite relative momentum. However,
the cross section is still enhanced by a factor three at threshold even for a
relative momentum of 
$p=450$ MeV, where the optical potential of the antikaon is already repulsive
(see Fig.\ \ref{fig:uopt_pkaon}). 
The temperature dependence of the cross
section is even stronger as anticipated from our discussion about temperature
effects for the antikaon optical potential.  
The peak structure vanishes already at a temperature of $T=30$ MeV. Still, an
enhancement factor of 2--3 remains at threshold even for the highest
temperature shown ($T=90$ MeV) for which the optical potential for the antikaon
is repulsive throughout the density region of interest (see Fig.\
\ref{fig:uopt_temp}).


\section{Effects from the selfenergy}
\label{sec:self}

A point of criticism can be raised in connection with the general approach. If
the $\Lambda(1405)$ as a resonance melts in hot and dense matter what should
remain is the mean-field potential due to the underlying antikaon-nucleon
potential.  
This can be tested by e.g.\ including an 
extra imaginary optical potential into the propagator which simulates
collisions with 
surrounding nucleons. The potential is known from proton-Nucleus scattering
data to be about $V_{NN}\approx -10$ MeV at $\rho_0$ (see e.g.\
\cite{Arnold81}).  Compared 
to the kaon optical potential inside the nucleus we expect a 10 \% effect if we
include the nucleon optical potential in the propagator (\ref{eq:propmat}).

The enhancement seen in the cross section originates from the dynamics of the
$\Lambda(1405)$ resonance. As we have demonstrated, it does not depend on the
detail of the interaction. Nevertheless, as recently shown by Lutz
\cite{Lutz98}, the $\Lambda(1405)$ does not move to higher energy, if the
propagator is calculated selfconsistently by including the self-energy of the
antikaons. The peak of the imaginary part of the isospin zero scattering
amplitude is not shifted but gets broader in the medium. Hence, strength is
still transported from the off-shell energy towards threshold energy as the
peak broadens.  Using the optical theorem, this means that the cross
section will still be enhanced at threshold in the medium. The enhancement in
the 
medium will be less pronounced as in our approach but we estimate from Fig.\ 4
of \cite{Lutz98} that it will be still a factor of four or so enhanced compared
to the vacuum.

Both selfenergies can be incorporated in our approach by modifying the
propagator (\ref{eq:propmat}) to
\begin{eqnarray}
g_1(E,k_f) &=& \frac{g^2}{\Lambda^2}
        \int_0^\Lambda \frac{d^3 p}{(2\pi)^3} \, \nonumber \\
&\times&        \frac{\Theta(k_f - | \vec{p}+M_N\vec{v}_\Lambda |)}
                {E - m_K - 
W_K(E-M_N-E_N(p),p_{CM}-p;\rho) - M_N - i V_{NN} - p^2/(2 \mu_{Kp})}
\label{eq:propmatmod}
\end{eqnarray} 
where $W_K$ stands for the optical potential of the kaon which
has to be calculated selfconsistently. Note that the optical potential has now
a real and an imaginary part, where the former part is defined by eq.\ 
(\ref{eq:uopt}) and the latter part correspondingly.  The optical potential $W_K$
is calculated iteratively and then put into the propagator in the next step.
This iteration scheme converges after 5--6 steps in agreement with the
calculations in \cite{Lutz98}. We denote the results of this approach in the
following as the selfconsistent ones.

\begin{figure}[htbp]
\begin{center}
\leavevmode
\psfig{file=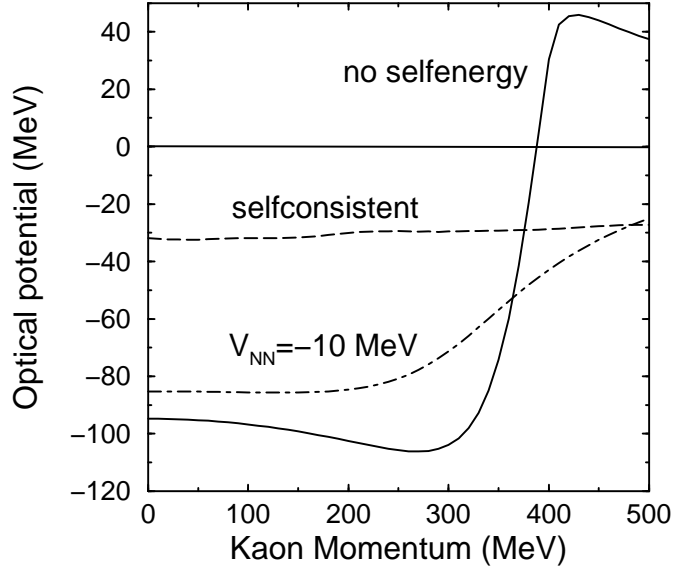,height=0.35\textheight} 
\end{center}
\caption{The optical potential including the selfenergy for the nucleons
(dash-dotted line) or antikaons (dashed line) as a function of the relative
momentum with respect to the nuclear matter rest frame.}
\label{fig:uopt_pkaon_comp}
\end{figure}

The effects of the nucleon and the antikaon optical potential on the 
real part of the antikaon optical potential are
summarized in Fig.\ \ref{fig:uopt_pkaon_comp}. Shown is the momentum dependence
at normal nuclear density. Our previous result without the implementation of
any selfenergy for the nucleon or the antikaon is plotted in solid.
At zero momentum, the optical potential changes by
10 \%, as expected, when including the nucleon optical potential. 
The selfconsistent approach reduces the optical potential remarkably to
$U_K(\rho_0) = -32$ MeV in stark contrast to the standard values deduced from
Kaonic atoms \cite{Fried94}. 
On the other hand, the momentum dependence of the
selfconsistently calculated optical potential is just flat. 
That means, that the optical potential does not turn repulsive anymore as it
happens for the calculation with the free propagator. The nucleon optical
potential also changes the momentum dependence considerably, so that the
potential decreases but is still attractive at high momenta. This trend
continues when applying values of $V_{NN}=-30,-50$ MeV: the optical potential
saturates at $-60$ MeV at zero relative momentum and 
at $-40$ MeV at high relative momentum.
Note, that for all cases considered here, the optical potential is still
substantially weakened at high momenta (or even over the whole momentum range
shown). We have also checked for a repulsive optical potential of the $\Sigma$ 
\cite{Batty94} of $U_\Sigma=+30$ MeV at $\rho_0$.
Choosing in addition $U_\pi=+30$ MeV and $U_\Lambda=-30$ MeV at $\rho_0$,
we find that the real part of the optical potential is reduced slightly by about
10\%. We also checked possible effects from an effective nucleon mass as done
in \cite{Koch94} before.
A nucleon mass shifted by $-50$ MeV at $\rho_0$ 
does not alter significantly the position and shape of the $\Lambda(1405)$
compared to the shifted threshold and hence can be safely neglected.
This result is in line with refs.\ \cite{Waas96b,Ramos99} where binding energy effects
for hyperons and nucleons have been found to be small, too.

\begin{figure}[htbp]
\begin{center}
\leavevmode
\psfig{file=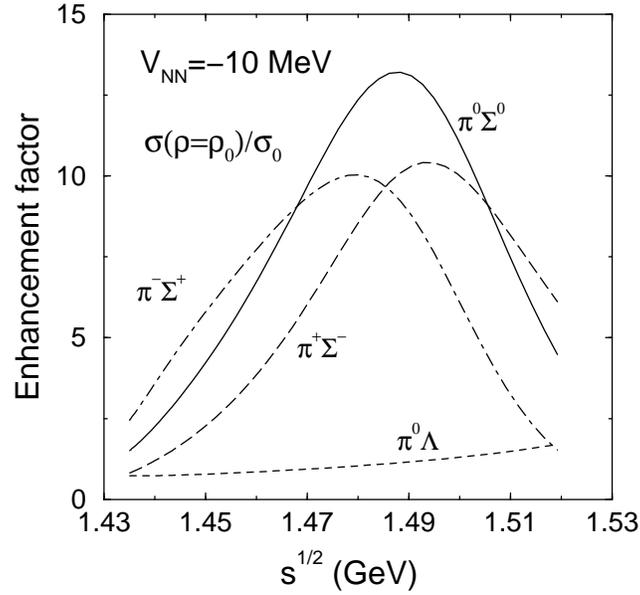,height=0.35\textheight} 
\end{center}
\caption{The enhancement factor of the antikaon production cross sections at
normal nuclear density including an imaginary potential for the nucleons.}
\label{fig:cross_nnpot_enh}
\end{figure}

\begin{figure}[htbp]
\begin{center}
\leavevmode
\psfig{file=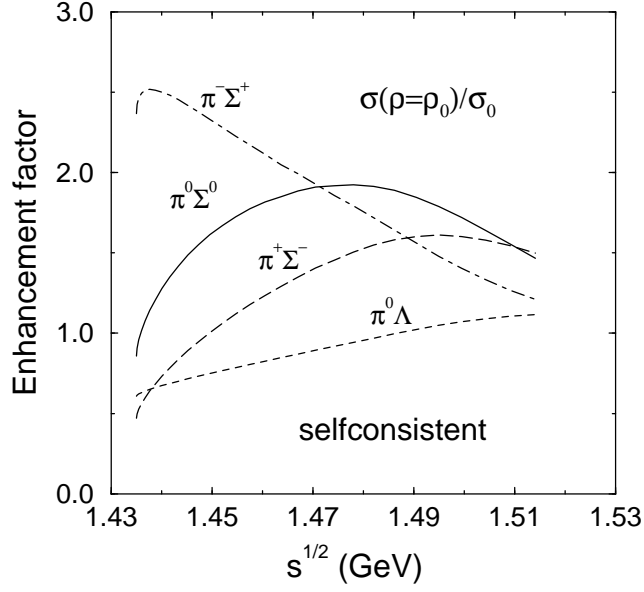,height=0.35\textheight} 
\end{center}
\caption{The enhancement factor of the antikaon production cross sections at
normal nuclear density for a selfconsistently calculated selfenergy of the
antikaon.} 
\label{fig:cross_nrselfc_enh}
\end{figure}

We turn now again to the production cross section of K$^-$ via hyperons, i.e.\
the enhancement factors when including the nucleon optical potential (Fig.\
\ref{fig:cross_nnpot_enh}) and for the selfconsistent calculation (Fig.\
\ref{fig:cross_nrselfc_enh}). The nucleon optical potential reduces the overall
enhancement factor by about a factor of two. The shapes are essentially the
same as for the free propagator, Fig.\ \ref{fig:cross_enh2}. For larger values 
of $V_{NN}=-30,-50$ MeV we find that the enhancement factors at the maximum
are reduced down to $4-5$ and $2.5-3$, respectively. Effects of hyperon and
pion optical potentials reduce the maximum enhancement factors down to
$16-22$, i.e.\ a repulsive $\Sigma$ potential still results in a sizable 
enhancement of the $\pi\Sigma$ production cross section.

The selfconsistent case shows a quite different behavior. The $\pi^0\Sigma^0$
and the $\pi^+\Sigma^-$ channels are enhanced at higher momenta of the outgoing
antikaon while the $\pi^-\Sigma^+$ channel increases especially at threshold up
to a factor of 2.5. Compared to the calculation done in \cite{Lutz98}, a
calculation using the chiral interaction terms provides a
larger enhancement. We attribute this to the energy dependence of the
chiral interaction term which will shift strength from lower to higher
energies.  


\section{Kaon Production in Heavy Ion Collisions with Modified Cross Sections}
\label{sec:production}
\subsection{Description of the model}
In order to explore the observable consequences of the results above we have
performed calculations of $K^-$ production in Ni+Ni collisions at 1.8 AGeV
within a BUU transport model that has been quite successfully applied to
particle production in heavy-ion collisions at SIS energies \cite{TeisZP97}
as well as pion \cite{Pion99} and photon \cite{Photo99,Cebaf99} 
induced reactions.
For a detailed description of the underlying formalism and 
the the non-strange part of the model we refer
to Refs.~\cite{TeisZP97,Photo99}. 
We note that for the calculations presented here the string fragmentation model
FRITIOF \cite{FRITIOF} is not included.
\subsubsection{Strangeness production}
Our model contains the following strangeness production channels in
baryon-baryon collisions:
\begin{eqnarray*}
B B &\to& N \Lambda K \\ 
B B &\to& N \Sigma K  \\
B B &\to& N N K \bar{K}.
\end{eqnarray*}  
The cross section for $B B \to N Y K$ is parameterized in the following way:
\begin{equation}
\sigma_{B B \to N Y K}=C \frac{1}{p_i \sqrt{s}} |{\cal M}_Y|^2 16 (2 \pi)^7 
\Phi_3,
\end{equation}
where $p_i$ is the cms momentum of the incoming particles and
$\sqrt{s}$ is the total cms energy. $\Phi_3$ is the 3-body phase space as,
for example given by Eq.~(35.11) in Ref.~\cite{PDG}. For the 
isospin coefficient $C$ we adopt the one-pion-exchange model of 
Ref.~\cite{Randrup80}. We obtain the following expression of 
Clebsch-Gordan coefficients:
\begin{equation}
C=
| \langle \frac{1}{2} 1 m_3 m_\pi| \frac{1}{2} 1 t_1 m_1 \rangle |^2
\sum_I g_Y^I 
| \langle t_2 1 m_2 m_\pi| t_2 1 I M \rangle |^2
| \langle t_Y \frac{1}{2} m_Y m_K| t_Y \frac{1}{2} I M \rangle |^2
+ (1 \leftrightarrow 2),
\end{equation}
where $t_1$ and $t_2$ are the total isospins of the incoming baryons and
$m_1$ and $m_2$ are their $z$-components. The $z$-component of the 
isospin of the outgoing nucleon is given by $m_3$. The coefficient $g_Y^I$
denotes the weight of the $Y K$-system with isospin $I$. For $\Lambda K$ we
simply have $g_\Lambda^{1/2}=1$, for $\Sigma K$ we use
$g_\Sigma^{1/2}=2 g_\Sigma^{3/2}=\frac{2}{3}$ in order to describe the
experimental data as will be shown below.
\par
\begin{figure}[htbp]
\begin{center}
\leavevmode
\psfig{file=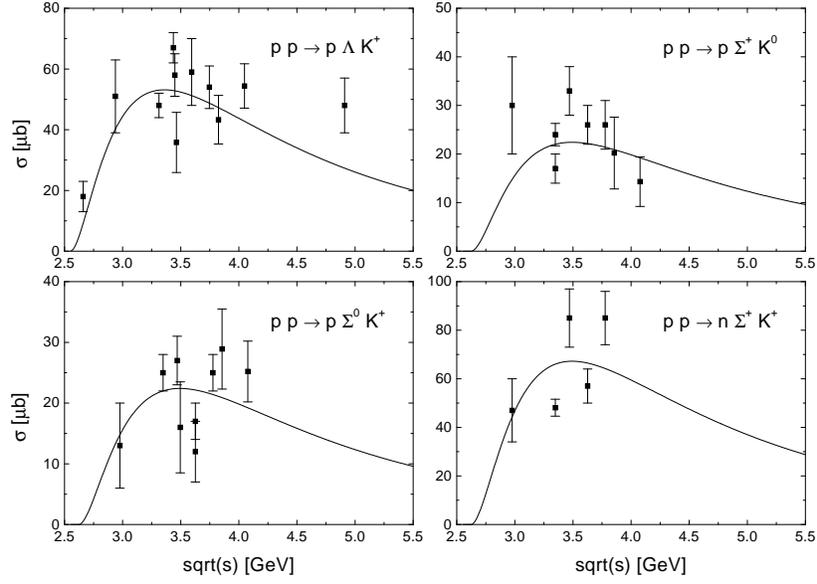,height=0.35\textheight} 
\end{center}
\caption{
Cross sections for $p p \to N Y K$. The experimental data are taken from
Ref.~\protect\cite{Landolt}.}
\label{fig:buu_pp}
\end{figure}

 The matrix element $|{\cal M}_Y|^2$ is parameterized as:
\begin{equation}
|{\cal M}_Y|^2=a_Y \left( \frac{1}{s/{\rm GeV}^2-3.94} \right)^{1.781},
\end{equation}
with $a_\Lambda=59.3$ mb and $a_\Sigma=44.5$ mb. In Fig.~\ref{fig:buu_pp}
we show that this gives a reasonable description of the experimental data
for proton-proton collisions. Unfortunately there are presently no
experimental data for proton-neutron collisions in the energy
range of interest available. We note that the one-pion-exchange model gives:
\[
\sigma_{p n \to N \Lambda K}=5 \sigma_{p p \to p \Lambda K^+},
\] 
because the model favors transitions with total isospin $I_{tot}=0$. This
might not be realistic. Under the assumption that the total matrix element
does not depend on the total isospin one gets
$\sigma_{p n}=\sigma_{p p}$ which is a factor of 5 smaller. 
For the process $B B \to N N K \bar{K}$ we use the parameterization given
in \cite{Cassing97}.  
\par In the meson-baryon sector our model includes the following processes
for strangeness production:
\begin{eqnarray*}
\pi N &\leftrightarrow& \Lambda K \\
\pi N &\leftrightarrow& \Sigma K \\
\pi N &\to& N K \bar{K} \\
\pi \Delta &\leftrightarrow& \Lambda K \\
\pi \Delta &\leftrightarrow& \Sigma K \\
\pi \Delta &\to& N K \bar{K}.
\end{eqnarray*}
The cross sections for $\pi N \to Y K$ and $\pi \Delta \to Y K$ 
are adopted from Ref.~\cite{Huang}. For $\pi N \to N K \bar{K}$ we use
the parameterization of Ref.~\cite{Sibirtsev}. For $\pi \Delta \to N K \bar{K}$
the same isospin averaged cross section is employed. The different 
isospin contributions are calculated using the Feynman diagram of
Ref.~\cite{Sibirtsev} with a $\Delta$ instead of a nucleon. 
The $\pi N$ and $\pi \Delta$ cross sections described here are also used 
for $\pi N^\star$ and $\pi \Delta^\star$ scattering.
\par In addition, our model includes the process:
\[ \pi \pi \to K \bar{K}, \]
where we adopt the isospin averaged cross section from \cite{Cassing97} and
calculate the isospin coefficients under the assumption that the matrix
element does not depend on the total isospin and the different contributions
add up incoherently.   
\subsubsection{Kaon-nucleon interaction}
In the $S=1$ channel our model contains the processes:
\begin{eqnarray*}
K N &\to& K N \\
K N &\to& K N \pi,
\end{eqnarray*}
where the used cross sections are described in detail in Ref.~\cite{Cebaf99}.
\par In the $S=-1$ channel we usually take into account the hyperon
resonances explicitly \cite{Cebaf99}. In the present work they are
not explicitly propagated in order to allow for an easier implementation
of the medium modifications described above. Without the explicit resonance
propagation their contributions are added to the 'background' cross sections.
The following processes are included:
\begin{eqnarray*}
\bar{K} N &\to& \bar{K} N \\
\bar{K} N &\leftrightarrow& \pi \Lambda \\
\bar{K} N &\leftrightarrow& \pi \Sigma \\
\bar{K} N &\to& \pi \pi \Lambda, \pi \pi \Sigma, \pi \bar{K} N.
\end{eqnarray*}
The used cross sections can be found in Ref.~\cite{Cebaf99}. We have checked
that an explicit inclusion of the hyperon resonances does not change our 
results. We have
also verified that the vacuum cross sections from the coupled channel
model described above agree very well with the cross sections that are
normally used in our transport model. For the present work we always use
for invariant energies $\sqrt{s}$ below 1.52 GeV the cross sections from
the coupled channel model, i.e. also for the calculations without
medium modifications. 
\subsection{Results} 
\begin{figure}[htbp]
\begin{center}
\leavevmode
\psfig{file=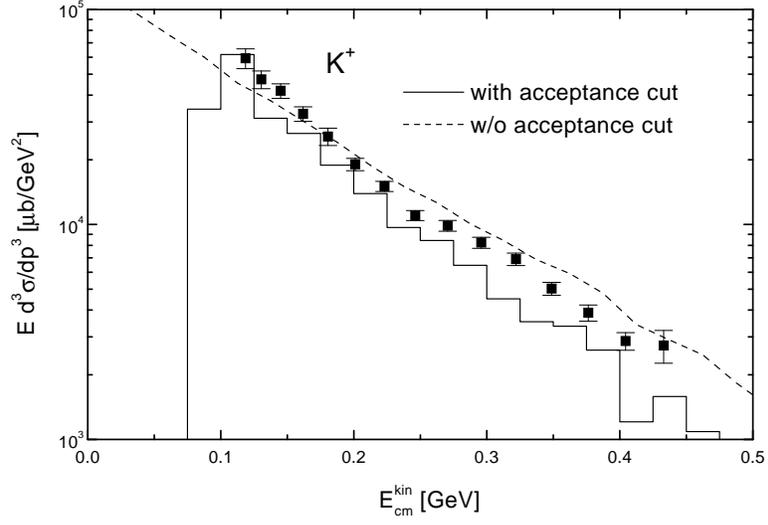,height=0.35\textheight} 
\end{center}
\caption{
Inclusive invariant $K^+$ production cross section for Ni+Ni collisions
at 1.8 AGeV. For the dashed line the experimental acceptance cut
($40^\circ < \theta_{lab} < 48^\circ$) was not taken into account. 
The experimental data are taken from
Ref.~\protect\cite{Kaos97}.
}
\label{fig:buu_kp}
\end{figure}

First, we note that our model gives a reasonable description of the 
experimental data on pion production and nucleonic observables in Ni+Ni 
collisions at 1.8 AGeV. In Fig.~\ref{fig:buu_kp} we compare our results for
the $K^+$ spectrum with the experimental data from Ref.~\cite{Kaos97}.
One sees that our calculation (solid line)
describes the experimental data reasonably
well although we underestimate the spectrum by about 30\%. We also
show the result of a calculation (dashed line)
for which we did not take into account
the experimental acceptance cut ($40^\circ < \theta_{lab} < 48^\circ$) for
the kaon. This curve agrees better with the data. Compared to the
calculation with acceptance cut the slope of the spectrum is larger. At
this place we should mention that the calculations in \cite{Li97,Cassing97}
obtain contributions to the spectrum for kaon center of mass 
energies below 50 MeV 
for which the experimental acceptance vanishes.    
\par
\begin{figure}[htbp]
\begin{center}
\leavevmode
\psfig{file=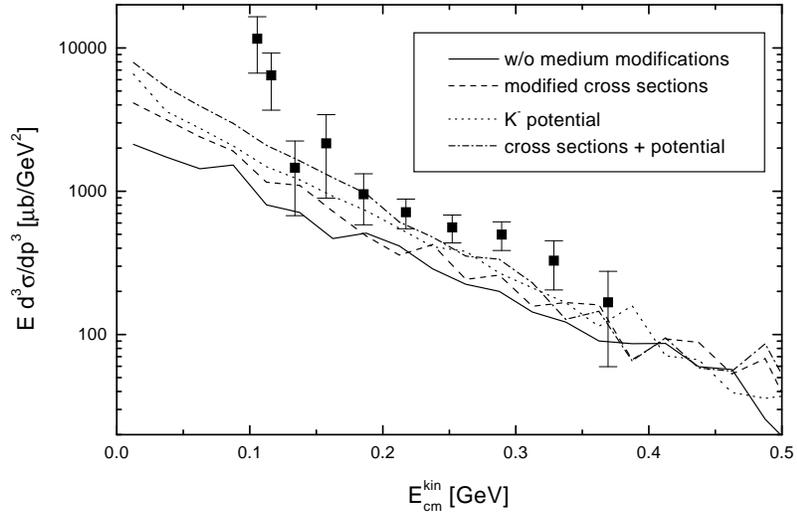,height=0.35\textheight} 
\end{center}
\caption{Inclusive invariant $K^-$ production cross section for 
Ni+Ni collisions
at 1.8 AGeV. Note that in the theoretical calculations the experimental
acceptance cut is not applied. The experimental data are taken from
Ref.~\protect\cite{Kaos97}.} 
\label{fig:buu1}
\end{figure}

 Having shown that our model gives a fair description of total strangeness
production we now turn to the investigation of observable
consequences of the above described medium modifications for the antikaon. 
Since the medium effects for the cross sections as well as the potential
are most pronounced at zero temperature and momentum we first present results
for which we neglect the temperature and momentum dependence and take all
quantities at $T=0$ and $p=0$
in order to explore the maximum possible effects.
Note that in all calculations \cite{Li97,Cassing97} the
temperature and the momentum dependence of the optical potential for
the antikaon have been neglected. 
In Fig.~\ref{fig:buu1} we compare our calculations to the
experimental data from Ref.~\cite{Kaos97}. Due to limited numerical statistics
we did not apply the experimental angular acceptance cut 
($40^\circ < \theta_{lab} < 48^\circ$) to our calculations. Because of this
we get contributions to the spectrum at low center of mass 
energies for which the 
experimental acceptance vanishes. We have checked that within the 
numerical fluctuations our results do not change when applying the cut
although 30\% effects like in the case of the $K^+$ can not be
excluded. 

\begin{figure}[htbp]
\begin{center}
\leavevmode
\psfig{file=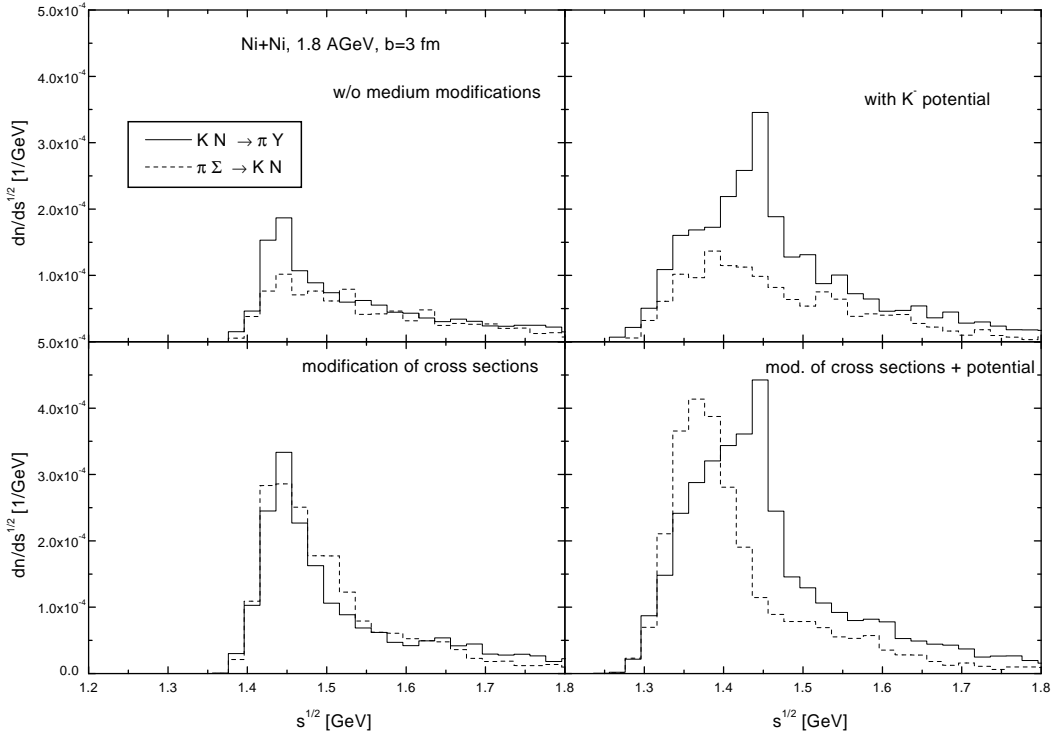,height=0.45\textheight} 
\end{center}
\caption{Time integrated numbers of $\pi \Sigma \to \bar{K} N$ and 
$\bar{K} N \to \pi Y$ collisions as function of $\sqrt{s}$ for different
medium modifications. The numbers were obtained from a calculation of a 
Ni+Ni collision at 1.8 AGeV and $b=3$ fm.} 
\label{fig:buu2}
\end{figure}

From Fig.~\ref{fig:buu1} one sees that the medium modification of the
cross sections (dashed line)
gives an enhancement of low energy $K^-$ by about a factor of
2 compared to the calculation without medium modifications (solid line)
but only a small effect for kinetic energies above 200 MeV. 
As stated above, the density dependent cross sections for $\pi Y$ and 
$\bar{K} N$ collisions from the coupled channel model has been implemented for 
invariant energies below 1.52 GeV (solid lines in 
Figs.~\ref{fig:cross_nr_pkaon} and \ref{fig:cross_nr_temp2}). 
Above that the usual vacuum cross sections have been used.
The effect of the potential alone (dotted line in Fig.~\ref{fig:buu1})
is quite similar.
A simultaneous application of modified cross sections and potential 
(dash-dotted line) leads to 
a result which gives an enhancement of low energy antikaons by about
a factor of three and a rather well description of the experimental data
except for the two lowest data points.
One should note that the modification of the 
$\pi Y \leftrightarrow \bar{K} N$ cross sections also gives an enhanced 
$\bar{K}$ absorption. In Fig.~\ref{fig:buu2} we therefore show the number
of $\pi \Sigma \to \bar{K} N$ (dashed lines) and $\bar{K} N \to \pi Y$ 
(solid lines) collisions for the
different scenarios. One sees that the modification of the cross sections
enhances the $\bar{K}$ production in $\pi \Sigma$ collisions for
invariant energies below 1.52 GeV by about a
factor of 3. 
An inclusion of the potential leads to an increase of $\bar{K} N \to \pi Y$
collisions by about a factor of two since more antikaons are overall produced,
especially in baryon-baryon and pion-baryon collisions, while the 
$\bar{K}$-production in $\pi \Sigma$-collisions is only slightly enhanced.
Due to the attractive $\bar{K}$-potential the $\sqrt{s}$-distributions
are shifted to lower energies.
In the calculation with modified cross sections and potential the
antikaon production in $\pi \Sigma$-collisions below 1.52 GeV is increased
by more than a factor 5 compared to the case without any medium
modifications. However, also the $\bar{K}$ absorption is substantially 
increased.

\begin{figure}[htbp]
\begin{center}
\leavevmode
\psfig{file=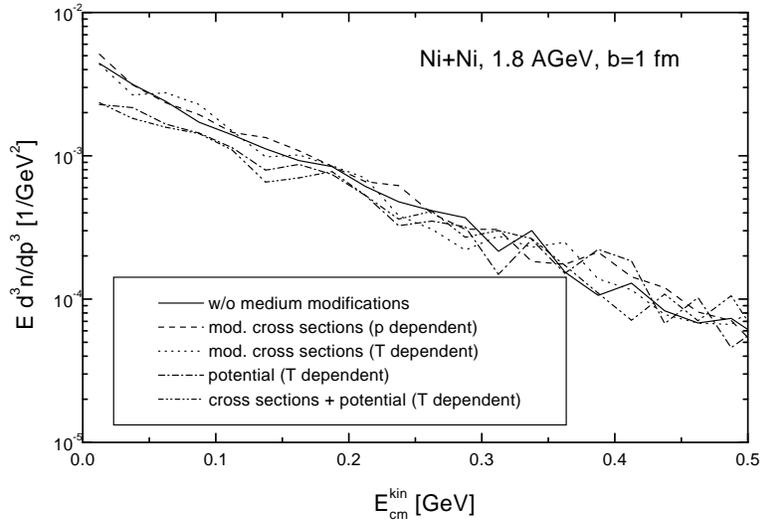,height=0.35\textheight} 
\end{center}
\caption{Effects of momentum or temperature dependent in-medium 
modifications on the $K^-$ spectrum in a Ni+Ni collision at 1.8 AGeV and
$b=1$ fm. Fluctuations are caused by low numerical statistics.} 
\label{fig:buu3}
\end{figure}

In Fig.~\ref{fig:buu3} we present the results of calculations for which
the momentum or temperature dependence in addition to the density dependence 
of the cross sections or potential were taken into account. 
Now the medium effects on the $K^-$ yield completely disappear because
-- as discussed above -- all modifications decrease with increasing momentum
and temperature. Using the temperature dependent $K^-$ potential from
Fig.~\ref{fig:uopt_temp} we even get a slight reduction of the $K^-$ yield
because in the dynamical calculation of the heavy-ion collision the
density-temperature correlations are such that the $K^-$ potential is almost
always repulsive. 

\begin{figure}[htbp]
\begin{center}
\leavevmode
\psfig{file=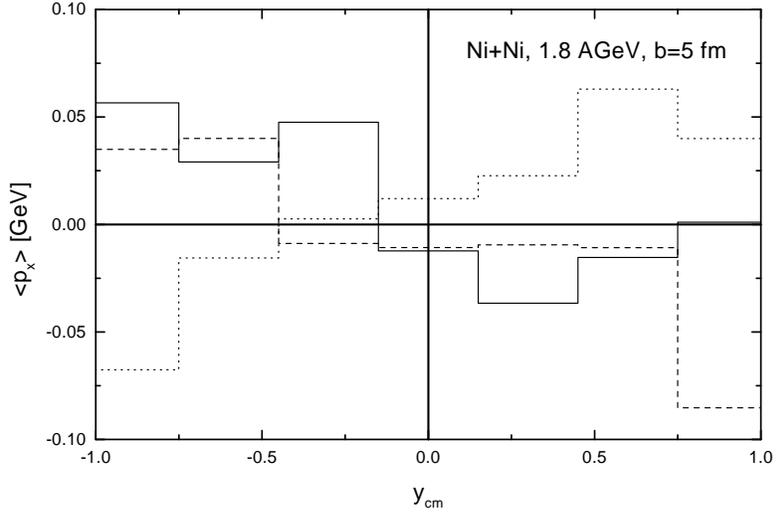,height=0.35\textheight} 
\end{center}
\caption{Effects of in-medium modifications on transverse $K^-$ flow in a 
Ni+Ni collision at 1.8 AGeV and $b=3$ fm: Without in-medium 
modifications (solid line), with modified cross sections for 
$\pi Y \leftrightarrow \bar{K} N$ at $p=T=0$ (dashed line), with the 
$\bar{K}$-potential at $p=T=0$ (dotted line).} 
\label{fig:buu4}
\end{figure}

One should note that a transport theoretical calculation of the 
total $K^-$ yield in heavy-ion collisions at SIS energies is encumbered 
with substantial uncertainties with respect to the input of unknown elementary
cross sections, like e.g. $\Delta \Delta \to \bar{K} X$ or 
$\pi \Delta \to \bar{K} X$. In this
context we again mention that even strangeness production in 
proton-neutron collisions at the relevant energies has not yet experimentally 
been measured. Therefore a determination of the $K^-$ potential from the
total yield appears to be difficult. 
An observable which might not be sensitive to these uncertainties is 
$K^-$ flow. Therefore, we show
in Fig.~\ref{fig:buu4} the influence of the different
in-medium scenarios on transverse $K^-$ flow. The anti-flow which is 
present in the calculation without medium modifications (solid line) 
is hardly         
affected by the modification of the cross sections (dashed line). 
The same also holds if
one includes the temperature or momentum dependence. The attractive $K^-$ 
potential (at $p=T=0$) (dotted line) is able to compensate this shadowing 
effect and
gives a positive $K^-$ flow since now the antikaons follow the nucleons.
An inclusion of the temperature dependence of the potential leads to
a disappearance of this effect.

\begin{figure}[htbp]
\begin{center}
\leavevmode
\psfig{file=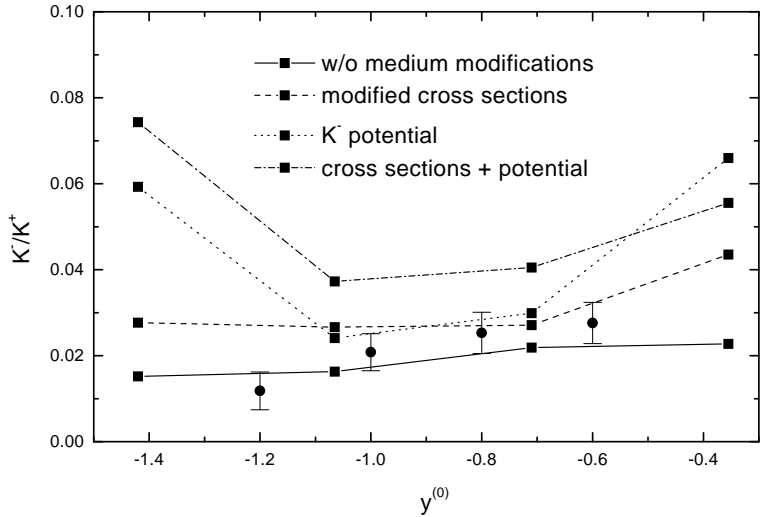,height=0.35\textheight} 
\end{center}
\caption{Effects of in-medium modifications on the $K^-/K^+$ ratio
in Ru+Ru collisions at 1.69 AGeV for $b<4$ fm as function of normalized
rapidity ($y^{(0)}=y_{beam}/y_{CMS}-1$):
Without in-medium modifications (solid line), with modified cross sections for 
$\pi Y \leftrightarrow \bar{K} N$ at $p=T=0$ (dashed line), with the 
$\bar{K}$-potential at $p=T=0$ (dotted line), with modified cross sections and
potential (dash-dotted line). The experimental data are taken from 
Ref.~\protect\cite{FOPI99}.} 
\label{fig:buu5}
\end{figure}

In Fig.~\ref{fig:buu5} we compare our calculations to the experimental data
of the FOPI collaboration \cite{FOPI99} for the $K^-/K^+$ ratio  
in $^{96}$Ru+$^{96}$Ru collisions at 1.69 AGeV. Here we have applied the
experimental acceptance cuts ($p_t>0.1$ GeV, 
$39^\circ < \theta_{lab} < 135^\circ$, $p_{lab}<0.32$ GeV). One sees that
already the calculation without medium modifications (solid line) 
gives a reasonable agreement with the experimental data although
the calculated distribution seems to be slightly too flat. The calculation 
with modified cross sections (at $p=T=0$) gives an enhancement of the 
$K^-/K^+$ ratio by about 50\% in the measured phase space region and is
also compatible with the experimental data. The 
attractive $\bar{K}$-potential (at $p=T=0$) (dotted line) leads to an
pronounced increase of the ratio towards mid-rapidity.
However, within the measured rapidity range also this scenario is hardly
distinguishable from the one without medium modifications. We note that the
theoretical calculation in the lowest rapidity bin around $y^{0}=-1.4$ suffers
from very low numerical statistics as the rapidity distributions are 
practically zero. 
The calculation with modified cross
sections and potential (dash-dotted line) gives a rather flat
distribution that overestimates the experimental data by about 50\%.
An inclusion of the temperature dependence again gives a result which is even
slightly below the calculation without medium modifications.

Our results can be compared with Refs.\ \cite{Li97,Cassing97,Li98}. 
The calculation for $K^-$ production in Ni+Ni collisions
at 1.8 AGeV (Fig.~\ref{fig:buu1})
without employing medium modifications agrees rather well with the
respective calculation in Ref.~\cite{Li97}. The calculation in 
Ref.~\cite{Cassing97} gives a result that is a factor of 2 smaller. The effect
of the attractive $K^-$ potential is weaker in our calculations than in
\cite{Li97,Cassing97}. While in our calculations the $K^-$ yield at
energies around 200 MeV is only enhanced by about 50\%, the calculations
in \cite{Li97} and \cite{Cassing97} obtain factors of 2 and 3, respectively.

In the calculation of the $K^-/K^+$ ratio in Ru+Ru collisions at 1.69 AGeV
in Ref.~\cite{Li98} it appears that the experimental cuts have not been taken
into account since there are contributions at midrapidity where the 
acceptance vanishes.    
Without in-medium modifications this calculation 
is far below the data 
points \cite{FOPI99}
while an attractive in-medium potential (at zero relative momentum)
shifts the curve on top of the data points. We note that our calculation
without experimental acceptance cut gives a ratio at midrapidity which
is a factor of 3-4 larger than the calculation in Ref.~\cite{Li98}.


\section{Summary and Discussion}
\label{sec:summary}

We have studied the optical potential for antikaons in the nuclear medium for
finite relative momenta and finite temperatures. We find that the optical
potential changes remarkably and turns repulsive at a relative momentum
comparable to the nucleon 
Fermi momentum and/or a finite temperature of a few tens of MeV. 
The strong momentum dependence is reduced when the hadron selfenergies are taken
into account selfconsistently but it results in a substantially shallower
antikaon potential.
Hence, antikaons produced in the central region of an heavy-ion collision are
likely to feel a substantially less attractive if not repulsive
potential. Therefore, it seems that contrary to
previous assumptions the antikaon potential at zero temperature and momentum
can not be reliably used to study in-medium effects in heavy-ion collisions.

On the other hand, we demonstrate that the cross section for producing
antikaons via $\Sigma$ hyperons is considerably enhanced in the nuclear
medium. This holds also for model calculations implementing chiral symmetry,
relativistic propagators, optical potentials for nucleons or for a
selfconsistent calculation of the antikaon selfenergy. The cross sections is always
enhanced at threshold, especially for $\pi^-\Sigma^+$, even for finite
relative momentum or finite temperature.

We implemented the in-medium effects for the antikaons within a BUU transport
model. An enhancement of the antikaon production is seen for medium
modified cross sections and a $K^-$ potential which follows more closely the
experimental data than the calculation without any of those in-medium effects.  
Including temperature effects, the in-medium enhancement vanishes and even
turns to a suppression of $K^-$ due to the repulsive potential felt by the $K^-$. 

Let us finally stress again that the environment for antikaons is completely
different in heavy-ion collisions compared to Kaonic atoms {\em and} neutron
stars where one probes the potential at zero momentum and zero temperature.
It appears that the enhanced subthreshold production of $K^-$ is more subtle
than hitherto assumed and deserves further study. Clearly, we do not have a
consistent picture to explain the enhancement seen experimentally. 
Further work has to be done to address this problem as for example by studying
in-medium effects using the chiral approach at finite momentum and temperature.
Future experimental data triggering on low momenta and/or the high density stage of
heavy-ion collisions might help to elucidate the possible in-medium effects for
antikaons in matter.

\acknowledgments

We are expressing our gratitudes to Gerry Brown for useful discussions and to
Avraham Gal for critical comments. 
We are grateful to N.~Herrmann for providing us with the
experimental data from the FOPI-collaboration. 
J. S.-B.\ thanks the nuclear theory group at LBNL, where part of this work was
completed, for their warm hospitality and thanks
the Alexander von Humboldt--Stiftung for their support during that time.
The work of M.E.\ and V.K.\ was supported by a NATO collaborative
research grant number 970102.
This work was supported partly by the BMBF, DFG, GSI, and the
Director, Office of
Energy Research, Office of High Energy and Nuclear Physics, Nuclear Physics
Division of the U.S. Department of Energy under Contract No. 
DE-AC03-76SF00098. 



\end{document}